\documentclass[sigconf]{acmart}

\AtBeginDocument{%
  \providecommand\BibTeX{{%
    \normalfont B\kern-0.5em{\scshape i\kern-0.25em b}\kern-0.8em\TeX}}}

\usepackage{array}
\usepackage{soul}
\usepackage{multirow}
\usepackage{xtab,booktabs,rotating}
\usepackage{flushend}
\usepackage{url}
\usepackage{balance}
\usepackage[colorinlistoftodos]{todonotes}
\usepackage{color, colortbl}
\usepackage[font=small,labelfont=bf]{caption}
\usepackage{algorithm,algpseudocode}
\usepackage{xspace}
\usepackage[inline]{enumitem}
\usepackage[T1]{fontenc}    
\usepackage{multirow}
\usepackage{mathtools}
\usepackage{caption}
\usepackage{subcaption}
\usepackage{empheq}
\DeclarePairedDelimiter\ceil{\lceil}{\rceil}

\setcopyright{acmcopyright}
\copyrightyear{2022}
\acmYear{2022}

\acmConference[TheWebConf '22]{TheWebConf '22: The Web Conference}{April 25--29, 2022}{TheWebConf, Fr}
\acmBooktitle{TheWebConf '22: The Web Conference,
 April 25--29, 2022, Lyon, Fr}

\newcommand{\sysname}{$\sf\small{Nebula}$\xspace}

\begin{document}

\title{\sysname:  Reliable Low-latency Video Transmission for Mobile Cloud Gaming}

\author{Ahmad ALHILAL}
\email{aalhilal@ust.hk}
\affiliation{%
  \institution{Hong Kong University of Science and Technology}
  \streetaddress{}
  \city{Hong Kong}
  \state{}
  \country{}
  \postcode{}
}
\author{Tristan BRAUD}
\email{braudt@ust.hk}
\affiliation{%
  \institution{Kong University of Science and Technology}
  \streetaddress{}
  \city{Hong Kong}
  \state{}
  \country{}
  \postcode{}
}
\author{Bo HAN}
\affiliation{%
  \institution{George Mason University}
  \streetaddress{}
  \city{}
  \country{USA}}
\email{bohan@gmu.us}

\author{Pan HUI}
\affiliation{%
  \institution{Hong Kong University of Science and Technology}
  \city{Hong Kong}
  \country{}
}
\email{panhui@cse.ust.hk}

\renewcommand{\shortauthors}{ALHILAL et al.}

\begin{abstract}
Mobile cloud gaming enables high-end games on constrained devices by streaming the game content from powerful servers through mobile networks. Mobile networks suffer from highly variable bandwidth, latency, and losses that affect the gaming experience. This paper introduces \sysname, an end-to-end cloud gaming framework to minimize the impact of network conditions on the user experience. \sysname relies on an end-to-end distortion model adapting the video source rate and the amount of frame-level redundancy based on the measured network conditions. 
As a result, it minimizes the motion-to-photon (MTP) latency while protecting the frames from losses.
We fully implement \sysname and evaluate its performance against the state of the art techniques and latest research in real-time mobile cloud gaming transmission on a physical testbed over emulated and real wireless networks. 
\sysname consistently balances MTP latency (<140\,ms) and visual quality (>31dB) even in highly variable environments. A user experiment confirms that \sysname maximizes the user experience with high perceived video quality, playability, and low user load.

\end{abstract}



\maketitle

\section{Introduction}

\begin{figure}[t]
    \centering
    \includegraphics[width=.8\linewidth]{ 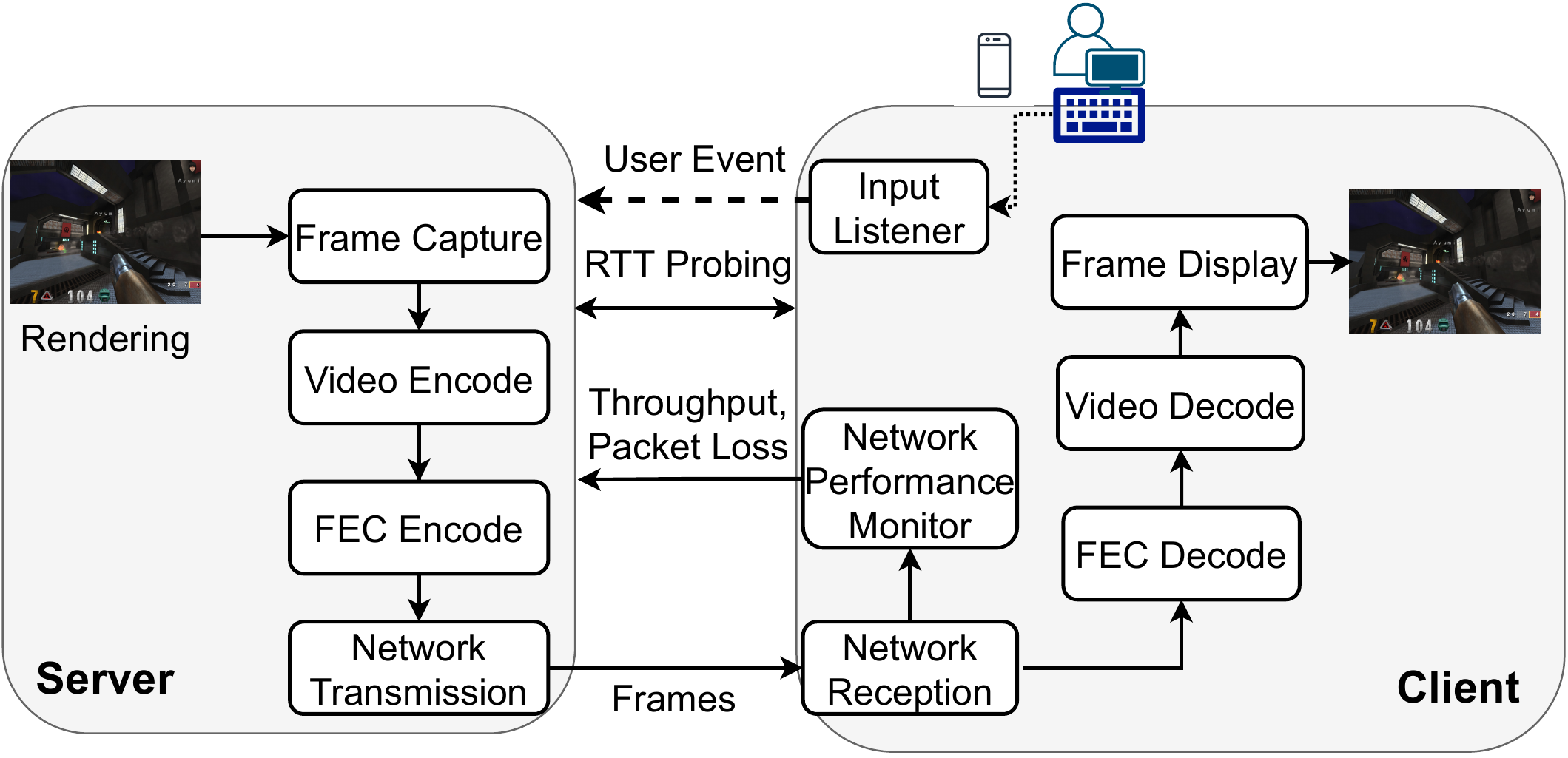}
    \caption{\sysname  minimizes latency while maximizing mobile cloud gaming video quality by performing joint source rate  and FEC redundancy control over unpredictable mobile wireless networks.}
    \label{fig:teaser}
\end{figure}

Cloud gaming enables high-quality video gaming on lightweight clients, supported by powerful servers in the cloud~\cite{wu2015enabling}.
As a highly interactive multimedia application, cloud gaming requires reliable and low-latency network communication~\cite{chen2011latencyCG}. 
Mobile cloud gaming (MCG) leverages the pervasivity of mobile networks to serve video game content on constrained mobile devices. 
Such networks exhibit unpredictable variations of bandwidth, latency, jitter, and packet losses, significantly impacting the transmission of game content.

Currently, most commercial cloud gaming platforms 
target sedentary gaming at home. 
These platforms often prioritize low latency over video quality through protocols such as WebRTC~\cite{carrascosa2020cloud,domenico2020network}.
Although WebRTC minimizes transmission latency, the resulting low video quality can have a detrimental effect on the users' quality of experience (QoE)\footnote{https://www.windowscentral.com/early-tests-call-google-stadias-picture-quality-question-and-thats-shame}.
Hence, improving the visual quality while preserving low latency remains a core challenge.
Mobile networks' inherent unpredictability further complicates the transmission of real-time MCG content. Sudden bandwidth drops reduce the amount of data that can be transmitted, 
while packet losses cause distortions in the decoded video flow. Cloud providers over public WANs present higher latency variation compared to providers with private network infrastructure~\cite{corneo2021surrounded}. Adaptive bit rate (ABR) and forward error correction (FEC) have been widely used to minimize the influence of bandwidth variations and packet loss on video transmission~\cite{perez2011effect}. However, most ABR solutions suffer from inaccurate estimations of the available bandwidth~\cite{8424813}, while FEC comes at the cost of high overhead. Besides, these two schemes have rarely been combined to optimize network usage and improve the QoE.

In this paper, we introduce \sysname, the first end-to-end framework combining \textit{video distortion propagation modelling} with \textit{adaptive frame-level FEC} and \textit{joint video encoding bitrate and FEC redundancy control}
for MCG over mobile networks (see Figure~\ref{fig:teaser}). 
\sysname relies on a mathematical model of video distortion that considers, for the first time, the propagation of errors among multiple game frames. As such, contrary to existing solutions that apply FEC to the entire group of picture (GoP)~\cite{wu2017streaming,wu2015leveraging},  \sysname applies frame-level FEC and unequally protects GoP frames prioritized by their proximity to the GoP's intra-frame. \sysname thus minimizes the end-to-end latency while significantly attenuating the effect of packet loss and the propagation of distortion in the GoP. \sysname also integrates source rate control to mitigate the effect of bandwidth variation.

We evaluate \sysname against standard TCP Cubic- and WebRTC-based video streaming solutions, and recent research on video streaming (Buffer-Occupancy, BO)~\cite{10.1145/2619239.2626296,huang2013bo} and mobile cloud gaming (ESCOT)~\cite{wu2017streaming}.
In both emulated environments and in-the-wild, \sysname balances motion-to-photon (MTP) latency and visual quality. TCP Cubic may present a higher visual quality, but the MTP latency shoots up with variable bandwidth. WebRTC keeps the MTP latency low at the cost of significant degradation of the visual quality, especially on networks with high jitter. BO and ESCOT critically affect video quality and latency, respectively.
A user study performed with 15 participants over a real-life WiFi network confirms the benefits of \sysname on the QoE. \sysname results in higher perceived video quality, playability, and user performance while minimising the task load compared to all other streaming solutions.


Our contribution is fourfold:
\begin{itemize}
    \item We propose the \textit{first model of end-to-end distortion} that accounts for the error propagation over a GoP.
    \item We introduce \textbf{\sysname}, an end-to-end framework for joint source/FEC rate control in MCG video streaming.
    \item We implement \sysname and the typical transmission protocols within a functional cloud gaming system.
    \item We \textit{evaluate \sysname} over a physical testbed through extensive experiments and a user study. \sysname  balances low-latency with resilience to distortion, leading to a good QoE.
\end{itemize}

\section{Background and Related Works}
\label{sec:works}

Although few works target MCG specifically, there is significant literature on real-time mobile video streaming. This section summarizes the major works on both video streaming and MCG systems.

\subsection{Real-time Mobile Video Streaming} 

FEC leads to high decoding delay when applied to full GoPs~\cite{wu2017streaming,wu2015leveraging}. Xiao et al.~\cite{xiao2013real} use randomized expanding Reed-Solomon (RS) code to recover frames using parity packets of received frames. They propose an FEC coding scheme to group GoPs as coding blocks, reducing the delay compared to GoP-based FEC~\cite{xiao2012supgop}. However, they do not address burst packet loss. Yu et al.~\cite{yu2014can} study the effects of burst loss and long delay on three popular mobile video call applications. The study highlights the importance of conservative video rate selection and FEC redundancy schemes for better video quality.
Frossard et al.~\cite{frossard2001jointmpeg2} adapt the source rate and FEC redundancy according to the network performance. Wu et al.~\cite{wu2013joint} propose flow rate allocation-based Joint source-channel coding to cope with burst and sporadic packet loss over heterogeneous wireless networks. 
Although these works employ FEC for video transmission, they do not address the low-latency requirements of MCG.

Congestion and overshooting are key sources of loss and distortion.  WebRTC controls congestion using Google Congestion Control (GCC) algorithm. GCC adapts the sending bitrate based on the delay at the receiver and the loss at the sender~\cite{jansen2018performance,carlucci2016analysis}. As a result, WebRTC favors real-time transmission to video quality. On the other hand, TCP Cubic~\cite{tcpcubic2008} is one the most widely deployed TCP variants in mobile networks that does not use latency as a congestion signal. TCP Cubic is sensitive to large buffers in cellular networks leading to bufferbloat~\cite{gettys2011bufferbloat}. 
The Buffer-Occupancy (BO) algorithm~\cite{10.1145/2619239.2626296,huang2013bo} selects the video rate based on the playback buffer occupancy. BO keeps a reservoir of frames with minimum quality and requests a lower rate upon low buffer occupancy signals. Such buffer-based technique encounters either low visual quality or lengthy latency, especially in extremely variable settings~\cite{10.1145/2619239.2626296}.

\subsection{Mobile Cloud Gaming} 
A few works target specifically MCG due to its novelty.
Provision of MCG is challenging owing to the high transmission rate and the limited resources. Chen et al.~\cite{chen2019framework} use collaborative rendering, progressive meshes, and 3D image warping to cope with provisioning issues and improve the visual quality. Guo et al.~\cite{guo2019qoe} propose a model to meet players' quality requirements by optimizing the cloud resource utilization with reasonable complexity. Outatime~\cite{lee2015outatime} renders speculative frames that are delivered one RTT ahead of time, offsetting up to 120ms of network latency. Wu et al.~\cite{wu2015hetnet} distribute the game video data based on estimated path quality and end-to-end distortion. In another work, they adjust the frame selection dynamically and protect GoP frames unequally using forward error correction (FEC) coding~\cite{wu2015enabling}. They refine the distortion model in~\cite{wu2017streaming} to support both source and channel distortion. Fu et al. ~\cite{fu2016distortion} investigate the source and channel distortion over LTE networks. All the above works apply GoP-level FEC, allowing them to cancel out error propagation at the cost of high motion to photon latency. In this work, we consider for the first time the distortion propagation across the GoP and adapt the frame-level FEC accordingly. As such, \sysname improves the quality of interaction and preserves the video quality beyond conventional (WebRTC and TCP Cubic) and research (ESCOT, BO) cloud gaming systems performance.

\section{Modelling the Joint source/FEC rate problem}
\label{sec:distortionmodel}

MCG relies on lossy video compression and protocols, leading to distortion in the decoded video. After justifying the need for joint source/FEC rate adaption, we refine existing distortion models to account for error propagation within a GoP, before formulating the joint source/FEC rate optimization problem. 

\subsection{Joint Source/FEC Rate Problem}
MCG is a highly interactive application that promises high-quality graphics on mobile devices. As such, a key metric of user experience is the MTP latency, defined as the delay between a user action and its consequences on the display. In mobile networks, ensuring a low MTP latency with a high visual quality is the primary challenge.

FEC minimizes the impact of packet losses and video distortion. However, the redundancy introduces significant overhead that increases latency and reduces the available bandwidth. Moreover, network congestion often leads to burst of losses~\cite{wu2017streaming}, for which FEC is often ineffective or even detrimental~\cite{yu2008model}. 
Although FEC
minimizes retransmission latency, most current works~\cite{wu2017streaming,wu2015enabling,wu2015leveraging} on FEC for cloud gaming apply FEC on the entire GoP, which either significantly increase the motion to photon delay (all GoP frames must be decoded before display), demand higher bandwidth, or decreases the visual quality. Besides, any large loss burst can prevent the recovery of the entire GoP.

We combine source rate with FEC redundancy control to mitigate congestion and residual packet losses and adapt the encoding bitrate of the video.
To minimize the MTP latency, we perform FEC at frame level instead of GoP level and vary the amount of redundancy based on the position of the frame in the GoP.

\subsection{Distortion Propagation Model}

End-to-end distortion represents the difference between the original frame at the sender, and the decoded frame at the receiver, and is represented by the Mean Square Error (MSE): 

\begin{equation}
MSE=\frac{1}{KMN}\sum\limits_{k=1}^{K}\sum\limits_{m=1}^{M}\sum\limits_{n=1}^{N}{\left(f_k(m, n)-\hat{f}_k(m, n)\right)}^{2}  
\end{equation}

\noindent where K is the number of frames with resolution MxN. $f_k(m, n)$ is the original pixel value and $\hat{f_k(m, n)}$ is the reconstructed pixel value at location $(m,n)$ of the $k^{th}$ frame.

We consider three distortion components:

\noindent\textbf{Encoder distortion $D_e$}: information lost during encoding. It provides the definitions for source distortion and transmission distortion as follows. Encoder distortion depends on the encoder rate  $R_e$ and empirical parameters $\alpha$ and $R_0$ so that $D_e = \frac{\theta_1 }{R_e - R_0}$~\cite{wu2015hetnet}. 

\noindent\textbf{Transmission distortion $D_c$}: caused by packet losses during network transmission. Depending on the  compression scheme, a packet loss may result into partial or complete frame loss. Transmission distortion is directly related to the packet loss rate $\Pi$ so that $D_c = \theta_2\Pi$ where $\theta_2$ is an empirical parameter~\cite{wu2015hetnet}.

\noindent\textbf{Decoder distortion $D_d$}: the additional distortion that propagates over multiple frames after a network packet loss. Existing works modelling distortion for FEC encoding applies FEC to the entire GoP so as not to consider error propagation in modelling video quality. However, in frame-level FEC encoding, losing any packet  corrupts the corresponding frame, directly impacting the subsequent frames in the same group of pictures (GoP). 
The model should thus account for such distortion propagation.

\begin{figure}[t]
    \centering
    \includegraphics[width=0.35\textwidth]{ 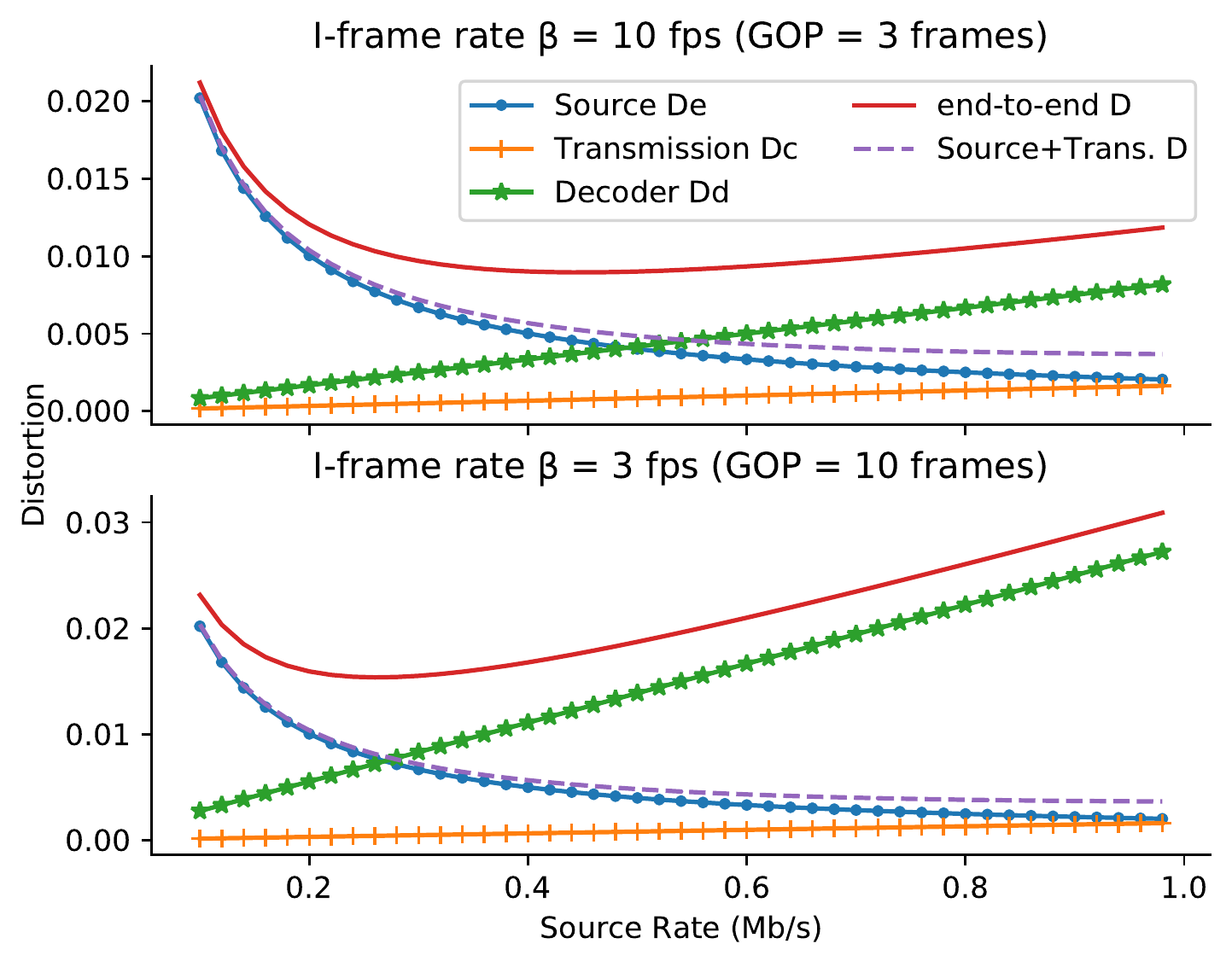}
    \caption{Source, Transmission, Decoder, and end-to-end distortion vs source video bitrate. Packet loss rate $\Pi=0.1\%$, UDP packet size $S=1500$B. When considering the interframe error propagation (decoder distortion), the end-to-end distortion increases with the source rate.}
    \label{fig:distortion}
\end{figure}

Let $\beta$ denotes I-frame rate, number of Intra-frames per second. We define  $D_d =  \frac{\theta_3\Pi}{\beta}$
where $\theta_3$ is an empirical parameter. 
The end-to-end distortion is the summation of the three components:
\begin{equation}
    D =  D_e + D_c + D_d = \frac{\theta_1 }{R_e - R_0} + \theta2 \Pi + \frac{\theta_3 \Pi}{\beta} 
    \label{eq:distortion}
\end{equation}

Consequently, the PSNR of the  pipeline is:
\begin{multline}
    PSNR=20 \log_{10}({MAX_I}) - 10  \log_{10}\left( \frac{\theta_1 }{R_e - R_0} + \left(\theta_2 + \frac{\theta_3}{\beta}\right)  \Pi  \right) 
\end{multline}

Figure~\ref{fig:distortion} displays the source, transmission, and decoder distortion, and their combined effect on the end-to-end distortion. We also display the end-to-end distortion as represented in previous models~\cite{wu2017streaming}. With our model, end-to-end distortion decreases at first, until the decoder distortion takes over. The end-to-end distortion then increases linearly with the bitrate. 
By considering for the first time interframe error propagation, our model raises a significant source of distortion for higher source rates.

\subsection{Joint source/FEC Coding Model}
\label{sec:jointmodel}

\begin{figure}[!t]
    \centering
    \includegraphics[width=0.8\columnwidth]{ 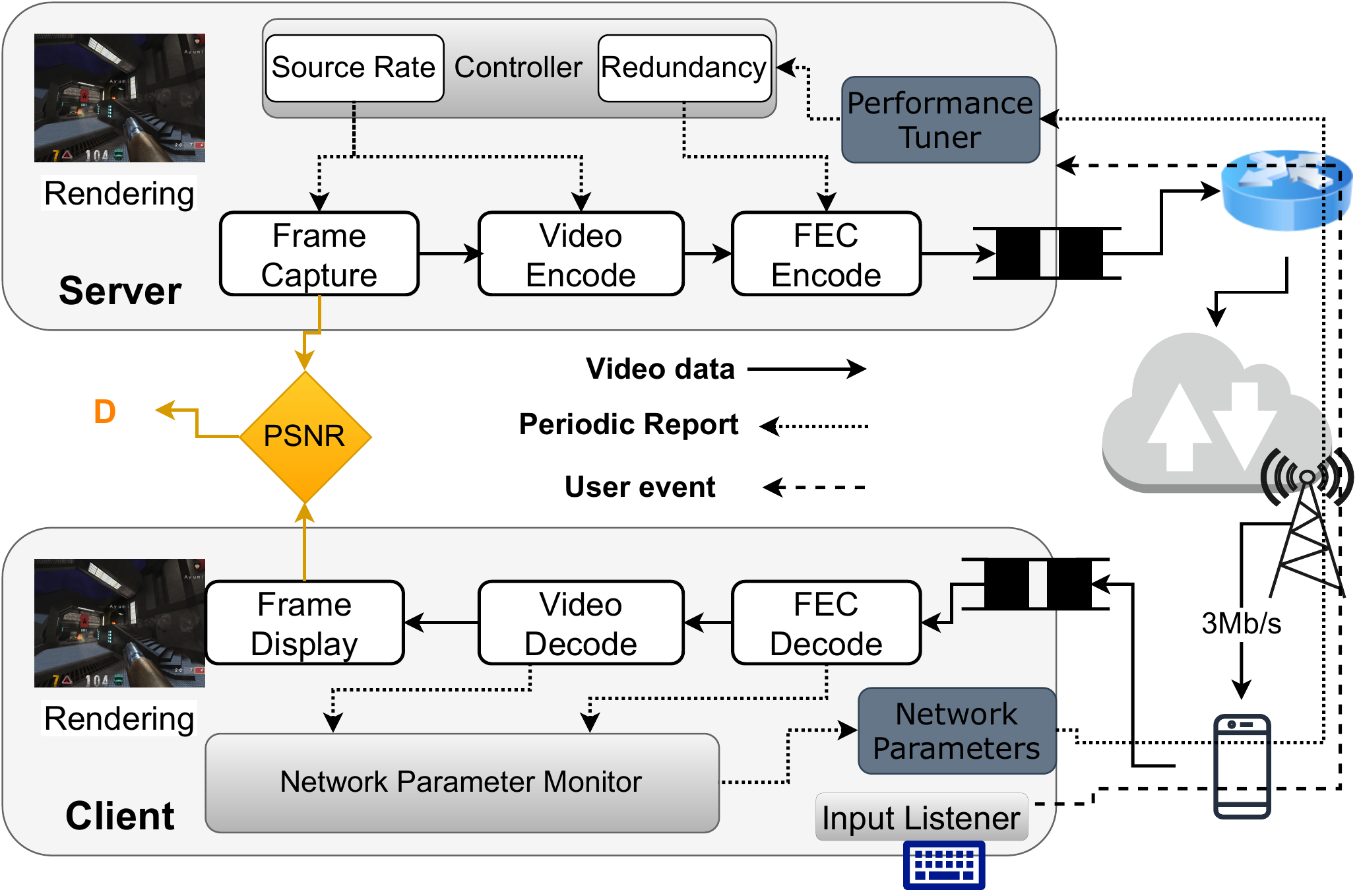}
    \caption{\sysname's detailed architecture. On the video data line, the server captures frames, video-encodes them into multiple spatiotemporal resolutions, FEC-encodes the output, and sends them to the clients. The client FEC-decodes, video-decodes the incoming stream into frames, and then displays them. On the feedback line, the client measures the network performance (Client$\rightarrow$Monitor), and sends the network parameters to the server so as to control the rate and redundancy (Server$\rightarrow$Controller). }
    \label{fig:sysarch}
\end{figure}

Distortion is a function of the available bandwidth and packet loss rate. 
The system objective is to minimize distortion while meeting the MCG constraints of real-time video transmission, limited and variable available bandwidth, and variable packet loss rate. 
We represent this objective as an optimization problem. 

The available bandwidth between the MGC server and client $\mu$ determines the rate constraint as the upper bound for the final sending rate ($R_e+R_r$). The goal of the adaption scheme is to find the optimal solution to minimize the end-to-end video distortion $D$, given the measured network parameters at the client side (i.e., RTT, $\mu$, $\Pi$ and MTP latency $MTP$), video encoding rate $R_e$ and delay constraint $T_d$. In MCG, interactive response  is critical to ensure a high quality of experience, with the MTP latency as the primary metric. The MTP latency is defined as the time between a user action in the game and its effect on the display. It comprises user input delivery, game execution, frame rendering and capturing, video encoding, transmission, and decoding, channel encoding, and playback delay. MTP is disproportional to both sending rate $R=R_e+R_r$ and available bandwidth $\mu$, and proportional to the queuing delay $Q_d=RTT-RTT_{min}$. The joint source and FEC coding adaptation problem for each frame can be formulated as follows.
\begin{equation}
\begin{aligned}
& \{R_e,R_r\} & =arg \min \quad D + \varphi MTP \\
& S.t: & D = \frac{\theta_1 }{R_e - R_0} + \theta_2  \Pi  + \frac{\theta_3\Pi}{\beta} \\
 &   &  MTP = \frac{\alpha_1}{\mu}+\frac{\alpha_2}{R_e+R_r}+\alpha_3Q_d + \alpha_4 \\
 &  \text{Latency Cnst } &  MTP \leq T_d \\
 &  \text{Throughput  Cnst } &  R_e + R_r  \leq \mu \\
\label{eq:optimDist1}
\end{aligned}
\end{equation}
where $\varphi$ is a hyper-parameter, $T_d$ is the upper-bound of MTP latency (i.e., 130ms), $\theta_1$, $R_0$, $\theta_2$ and $\theta_3$ are empirical parameters based on the encoder's configuration, and $\alpha_1$, $\alpha_2$, $\alpha_3$, and $\alpha_4$ are parameters derived from multivariate regression with goodness of fit $R^2=95\%$. In the regression, we use a collective dataset of 12 experiments over both wired and wifi networks to fit the linear model. Given the number of source packets $k$ and $\Pi$, we derive the  total number of packets (sum of source and redundant packets) $n$ as follows:\

\[
 n= 
\begin{cases}
    \Pi(t)>0,  \smash{\begin{cases} 
    max \left ( k+1,  \ceil{k \cdot (1 + \omega (F-f)\cdot \Pi(t) )} \right),  \hspace{1em}\texttt{cut $D_d$}\\
    max(k+1,  \ceil{(k\cdot(1+\Pi(t)) )},  \hspace{4em}\texttt{Otherwise}\\
    \end{cases} }\\\\
    \Pi(t)=0,   \hspace{1em} k  \tag{5}
    \label{eq:fecrate}
\end{cases}
\]
As the I-frame rate $\beta$ decreases, the decoding distortion, and thus the end-to-end distortion, increases rapidly as shown in Figure~\ref{fig:distortion}. The longer the error propagates, the higher the decoding distortion at the client side. To overcome the inter-frame error propagation that causes the decoding distortion, we protect the frames unevenly based on the frame index $f$ within the GoP. The redundant packets $k \cdot \omega \cdot \Pi \cdot (F-f)$ ensures higher protection as frame gets closer to the I-Frame (smaller $f$), where $0<\omega<0.4$ is an empirical weight that increases or decreases with the packet loss rate. When $\omega \cdot (F-f) = 1$, all frames are protected equally, and there is no extra protection against error propagation.

\subsection{Heuristic Model}
\label{sec:heuristic}
Minimizing the encoding distortion $D_e$ requires maximizing the encoding bitrate $R_e$. We do so heuristically by taking the maximum feasible source bitrate and the minimum redundancy rate at each time step $t$ as follows:

\[
 \begin{cases}
        Max ( R_e )  =  
\sum\limits_{t=0}^{T} Max( R_e(t) )\\
 Min ( R_r )  =  
\sum\limits_{t=0}^{T} Min( R_r(t) ), \vee t \in T \tag{6}
    \label{eq:maxrate}
\end{cases}
\]

To satisfy the latency and throughput constraints in the optimization problem~\ref{eq:optimDist1}, the sending bitrate ($R= R_e+R_r$) decreases automatically once experiencing a high round trip while not exceeding the measured throughput $\mu$ at the client. At a time t, the sending rate $R$ is upper-bounded by $\mu(t)*(1- Q_d(t))$, where $Q_d(t)=RTT(t)-RTT_{min}$ is the queuing delay, the difference between recent and min round trip time. Experimentally, satisfying the latency and throughput constraints prevents overshooting, thus satisfying the loss constraint implicitly. The formula that determines the sending bitrate $R$ is therefore defined as follows:

\[
 R= 
\begin{cases}
    \max  R_e(t) + \min R_r(t)   < \mu(t) (1 - Q_d(t) ),\\ \hspace{13em}  1 - Q_d(t)  >0 \\
    \min   R_e(t) +  \min R_r(t), \hspace{3em} 1 - Q_d(t) <= 0 \tag{7}
    \label{eq:sendrate}
\end{cases}
\]

With $t \in T$ is the current time,  and $T$ is the total streaming duration. 
$t$ takes discrete values. In our evaluation, we choose a time interval of 1 second for a 1-minute long video game session.

The system periodically reacts to the loss rate $\Pi$, available bandwidth $\mu$, and the MTP latency $MTP$ by adapting $R_e$ and $R_r$, avoiding overshooting and distortions that result from sporadic packet loss. 
\section{Distortion Minimization Framework}
\label{sec:sysmodel}

Figure~\ref{fig:sysarch} illustrates our proposed end-to-end framework. 
This framework aims to provide MCG systems with error resiliency and achieve optimal video quality by combining source rate control and FEC coding adaptation. After providing a general description of the framework architecture, we will focus on how the adaptive FEC coding and the adaptive source coding models operate.

\subsection{Framework Architecture}
The framework includes two components, a controller and a parameter tuner. 
The \textbf{parameter tuner} receives feedback from the client at runtime, tunes parameters related to the source and FEC encoding rate, and passes them to the controller.
The \textbf{controller} combines the video rate controller (to control the video encoder according to the tuned source coding rate $R_e$) and the redundancy rate controller (to control the RLNC coder according to the tuned redundancy rate $R_r$ and packet size $S$). Each captured frame at the server-side is compressed using the VP8 codec\footnote{\url{https://www.webmproject.org/code/}} and split into packets encoded for network transmission using RLNC~\cite{heide2018random}. The RLNC packets are transmitted using a UDP socket to the client's mobile device. The client decodes the RLNC packets to recover the frame data, which is then decoded to recover and display the game frame. 

The framework also includes the \textbf{network performance monitor} at the client-side to monitor the bandwidth, the loss rate,  the network round-trip-time, and the total MTP latency. These parameters are sent periodically to the parameter tuner at the server-side.

\subsection{Adaptive FEC Coder Model}

\begin{figure}[!t]
    \centering

    \includegraphics[clip,width=.4\textwidth]{ 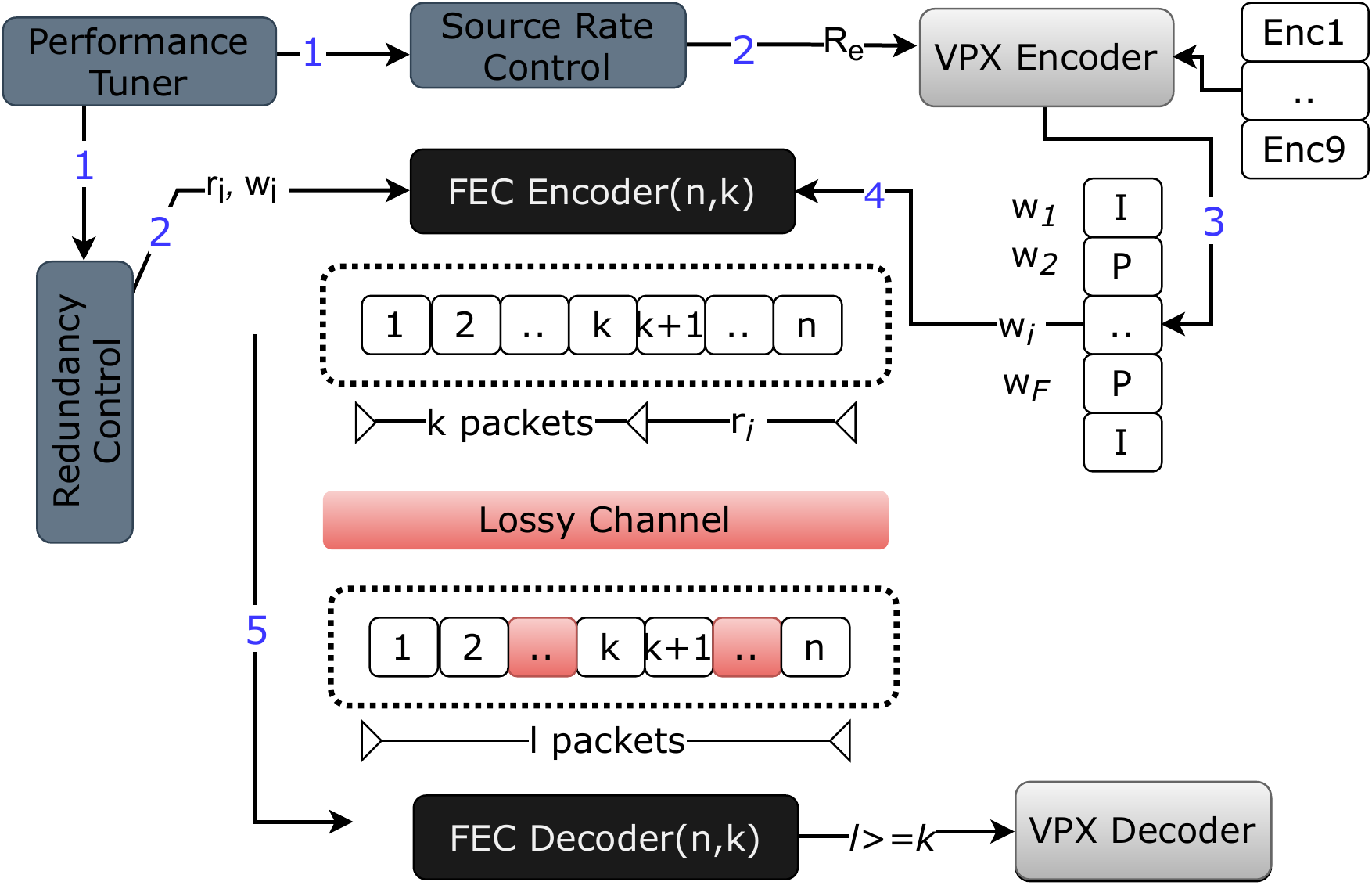}
        \caption{Illustration of VPX and FEC coding. VPX encoder uses multiple encoders (Enc1-9) for multiple bitrates (resolutions). Performance Tuner determines encoding and redundancy rate. FEC encoder applies unequal redundancy for GoP frames (equation~\ref{eq:fecrate}). FEC coder start decoding once $l\geq k$ packets arrived}\label{fig:vpxfeccoder}

\end{figure}

Our system uses Random Linear Network Coding (RLNC) to protect the game frames against channel loss. The frame data is divided into $k$ packets of size $S$ and linearly combined. The RLNC encoder generates a redundancy rate $R_r$ of $r=n-k$ redundant packets with an FEC block of size $n$ data packets~\cite{heide2018random}. The decoder recovers the frame if any random $k$ packets in the block are received, as shown in Figure~\ref{fig:vpxfeccoder}. The parameter tuner extracts the packet loss information from the client's report and tunes the redundancy based on the frame index within the GoP (see Equation~\ref{eq:fecrate}).
The variable FEC coding of the GoP frames  ensures lower error propagation as the frames leading to higher propagation receive  higher protection. 

\subsection{Adaptive Source Coding Model}

The bandwidth often shrinks when congestion occurs, leading to burst traffic loss~\cite{wu2017streaming}. 
Injecting more FEC-based redundant packets is ineffective to cope with congestion losses in the presence of burstiness~\cite{yu2008model}. 
As such, we introduce the source rate controller to control both the frequency of capturing the rendered frames and video encoding rate $R_e$. 
The parameter tuner extracts the available bandwidth and $MTP$ latency from the client's report adjusts $R_e$ by scaling down the game frames. It adapts to the client's bandwidth while minimizing both the distortion (see equation~\ref{eq:optimDist1}) and $MTP$ latency and relieving the congestion. Figure~\ref{fig:vpxfeccoder} illustrates the operation of the source rate control and redundancy control. By encoding videos several resolutions, the video encoder can dynamically adapt the bitrate of the video to the channel conditions. The video is encoded into nine  resolutions: HD (1080p, 720p), High (480p, 540p), Med (360p, 376p), and Low (270p, 288p and 144p), corresponding to bitrates of (6.5 Mb/s, 4.5 Mb/s), (3 Mb/s, 2 Mb/s), (1.8 Mb/s, 1.2 Mb/s), and (1 Mb/s, 0.6 Mb/s and 0.2 Mb/s), respectively. The source rate controller selects the closest bitrate to the measured throughput $\mu$. The resulting video dynamically changes resolution with minimal distortion as few packets are lost due to the bandwidth variations.

\section{Implementation}
\label{sec:implem}

We implement the distortion minimization framework into a functional prototype system. The framework leaves multiple aspects at the discretion of the developer, that we discuss in this section. 



\subsection{Video and Network codecs}

The framework is designed to be video and FEC codec-agnostic. However, we chose to focus on VP8 for video encoding/decoding and RLNC for FEC in our implementation.

\noindent\textbf{Video codec:} We use VP8~\cite{bankoski2011vp8}, an open-source codec developed by Google. Like H.264, VP8 compresses the game's frames  spatially (intra-frame coding) and temporally (inter-frame coding). VP8 presents it presents similar performance as H.264 (encoding time, video quality) while being open-source. VP8 also combines high compression efficiency and low decoding complexity~\cite{bankoski2011vp8tech}. We rely on libvpx, the reference software implementation of VP8 developed by Google and the Alliance for Open Media (AOMedia)~\footnote{https://www.webmproject.org/code/}. 

\noindent\textbf{FEC codec:}
We use Random Linear Network Coding (RLNC)~\footnote{\url{http://docs.steinwurf.com/nc_intro.html}} for channel coding. RLNC relies on simple linear algebraic operations with random generation of the linear coefficients. As such, RLNC is operable without a complex structure. RLNC can generate coded data units from any two or more coded or uncoded data units locally and on the fly~\cite{fouli2018rlnc}. 
The RLNC decoder can reconstruct the source data packets upon receiving at least $l$ linearly independent packets (see Figure~\ref{fig:vpxfeccoder}). To use RLNC encoding and decoding functionalities, we use Kodo~\cite{pedersen2011kodo}, a  c++ library for network/channel coding. 

\subsection{Client}

\textbf{Network Performance Monitor (NPM):}
This component measures the experienced packet loss rate $\Pi$, bandwidth $\mu$, round trip time $RTT$, and MTP latency ($MTP$). The link throughput $\mu$ is computed every $\Delta t$ as the net summation of the received frames' size over the net summation of frames' receiving time as follows:
\begin{equation} \tag{9}
   \mu =\frac{\sum^{F_s}_{f=1} \left( \sum_{p=2}^{k} S(p)\right) } {\sum^{F_s}_{f=1} t(f) }
    \label{eq:measuremu}
\end{equation}
where $F_s$ is the number of frames received in the current second, $f$ is the current frame of $k$ packets, $S(f)$ is the size of received frame, and $t(f)$ is the elapsed time to receive it.  
We start the frame's timer upon the reception of the frame's first packet, and disregard the size of this packet in $S(f)$. $\Pi$ is the ratio of the number of missed packets to $k$, the number of missing sequence numbers in the received packets until the successful recovery of the frame. $MTP$ is computed as the difference between the time of user input and the time for the corresponding frame to be displayed on the client screen. The NPM smoothes these values by applying a moving average over the latest five measurements.

\subsection{Server}
The server measures the $RTT$ through periodic \textbf{packet probing}~\cite{luckie2001towards}, at regular intervals  $\Delta t$.

\noindent\textbf{Parameter Tuner} tunes the source rate $R_e$ and the redundancy rate $R_r$ according to the packet loss rate $\Pi$, and network throughput $\mu$ received from the clients, and $RTT$. It adjusts the number of redundant packets $r=n-k$ according to the number of source packets $k$ and $\Pi$ based on equation~\ref{eq:fecrate}. It also determines the suitable spatial resolution of the game frames and frame capturing frequency to maximize quality with a source rate $R_e$ lower than $\mu$.

\noindent\textbf{Redundancy Controller} reads $R_r$ from the Parameter Tuner and updates the redundant packets for the  encoded video frame. Likewise, the \textbf{Rate Controller} reads $R_e$ to update the frame resolution.

\subsection{Prototype System Implementation}
\label{sec:pipelineimplem}

We implement our system as a python program with C and C++ bindings. We adopt a multi-process approach to improve performance and prevent a delayed operation to interrupt the entire pipeline. We 
separate the pipeline into 3 processes (frame capture, VP8 encoding, and RLNC encoding and frame transmission) at the server and 3 processes (frame reception and RLNC decoding, VP8 decoding, display) at the client. 
These processes intercommunicate using managed queues, where Queue and Process are classes in \texttt{multiprocessing\footnote{\url{https://docs.python.org/3/library/multiprocessing.html}}} module. 
We develop a protocol on top of UDP to carry server-to-client video data and client-to-server information.
Referring to figure~\ref{fig:sysarch}, the first server process captures the frames using the \texttt{mss} library. 
The video encoding process VP8-encodes them using a custom-written wrapper to the  libvpx\footnote{\url{https://github.com/webmproject/libvpx}}. Libvpx does not currently provide a GPU implementation, thus, encoding/decoding is performed on machines' CPU.
We use the python bindings of the \texttt{Kodo\footnote{\url{https://www.steinwurf.com/products/kodo.html}}} library to handle the RLNC operations. 
The server transmits packets carrying the encoded data to the client using FEC Forward Recovery Realtime Transport Protocol (FRTP) packets. 
The client recovers the encoded frame using the Kodo RLNC decoder. The VP8 video frames are decoded by the libvpx and displayed. The client reports the experienced network conditions periodically using the Network Performance/Parameters Report (NPR) packets. The client also listens to the user's input and sends the event information to the server. The server returns the event sequence number in the FRTP packet carrying the corresponding encoded frame, allowing the client to measure MTP latency. The server also probes the round-trip time periodically using RTTP packets (see Supplementary Materials - Figure~\ref{fig:nebulapackets}).

\subsection{Feasibility as a Web Service}

Many implementation choices in this section are consistent with the current practice in WebRTC.
Although most current web browsers provide VP8 encoding and decoding, they do not expose these functions to web applications. Instead, they provide an interface to the main functions of the libwebrtc. Similarly, we implement the core functions of \sysname as a C++ library to easily interface with web browsers with minimal logic in Python. \sysname is implemented primarily under an asynchronous fashion, exposing callbacks to applications that may use it. It can thus easily be implemented in current web browsers, and its API endpoints can be called through JavaScript with minimal modifications.

\section{Evaluation}
In this section, we evaluate \sysname through both system and user experiments. We first evaluate the system in terms of objective latency (MTP) and visual quality (PSNR) measures, before inviting users to rate their experience based on subjective metrics. All experiments are performed over a controlled testbed, with both controlled and uncontrolled network conditions to ensure reproducibility. 


\label{sec:eval}

\subsection{Evaluation Setup}
\label{sec:testbed}

\begin{figure}[t]
    \centering
    \includegraphics[width=\linewidth]{ 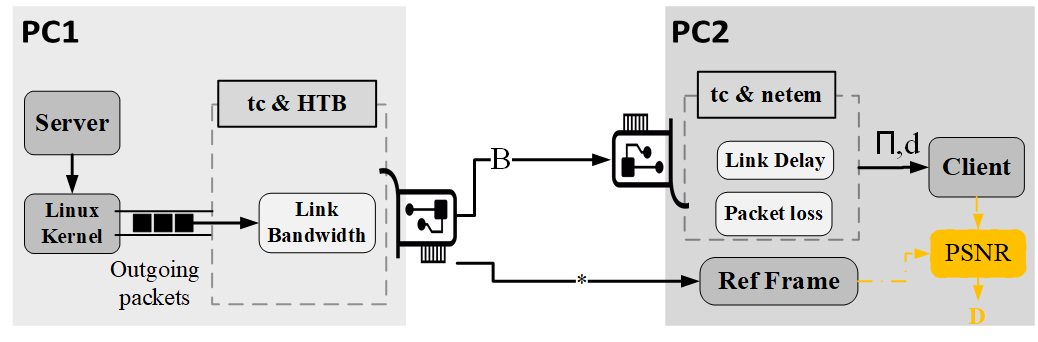}
    \caption{Experimental Testbed. It emulates a mobile network's link bandwidth, delay and packet loss using tc, netem and HTB. A reference link transmits original frame to compute the distortion $D$.}
    \label{fig:testbed}
\end{figure}

We characterize \sysname over the physical testbed represented in Figure~\ref{fig:testbed}. This testbed is composed of two computers: PC1 as a server (Ubuntu 18.04 LTS, Intel i7-5820K CPU @ 3.30GHz) and PC2 as a client (Ubuntu 18.04 LTS, Intel(R) Core(TM) i5-4590 CPU @ 3.30GHz). Using a computer as a client simplifies the implementation of the prototype system and the system measures. The two computers are  connected first through an Ethernet link to and ensure control  and reproducibility. On PC1, we emulate a wireless link with bandwidth $\mu$, latency $d$, and packet loss rate $\Pi$ using \texttt{tc}\footnote{\url{https://man7.org/linux/man-pages/man8/tc.8.html}}, \texttt{NetEm}\footnote{\url{https://man7.org/linux/man-pages/man8/tc-netem.8.html}}, and Hierarchy Token Bucket (HTB)\footnote{\url{https://linux.die.net/man/8/tc-htb}}. We then connect the computers through the eduroam WiFi to perform in-the-wild experiments. On both machines, we set up the kernel HZ parameter to 1,000 to avoid burstiness in the emulated link up to 12Mb/s.

On this testbed, we compare \sysname to the following baselines (see Table~\ref{tab:baselines} in~\ref{sec:appendix}): TCP Cubic, WebRTC, Buffer Occupancy (BO)~\cite{10.1145/2619239.2626296,huang2013bo}, and ESCOT~\cite{wu2017streaming}. These baselines are  standard in network transmission (TCP Cubic), in video streaming (WebRTC), or correspond to the latest related research works (BO, ESCOT). We implement all baselines in the pipeline described in Section~\ref{sec:pipeline} as follows. TCP Cubic is provided by the Linux Kernel. We integrate WebRTC  using \texttt{AioRTC}\footnote{\url{https://github.com/aiortc/aiortc}}, that we modify to support customized source streams at the server and remote stream track of WebRTC at the client (see  Supplementary Materials - Figure~\ref{fig:baselinesfw}). We  implement BO as a queue-based playback buffer in the pipeline. Finally, we modify \sysname to perform FEC at GoP level to implement ESCOT. We implement a user-event listener to send user's input to the server. The server executes the game logic and renders the frames upon event arrival. The architecture of the WebRTC implementation requires us to measure the visual quality in real-time. As such, we use the PSNR as it is the least computationally-intensive measure. We monitor the network conditions and all performance metrics under the same method for all experiments to ensure results consistency.

\subsection{Pipeline Characterization}
\label{sec:pipeline}

\begin{figure}[!t]
    \centering
    \includegraphics[width=.75\linewidth]{ 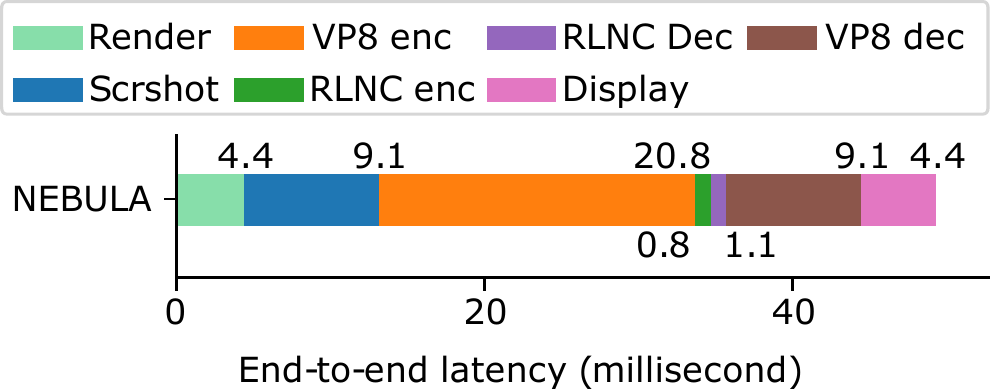}
    \caption{Latency Decomposition of the MCG Pipeline (without the additional network latency). }
    \label{fig:latency}
\end{figure}

We first evaluate the prototype system's pipeline presented in Section~\ref{sec:pipelineimplem}. We set up the testbed with a fixed bandwidth of 20\,Mb/s, latency of 50\,ms, and PLR of 1\%. 
These network conditions correspond to a typical client-to-cloud link~\cite{ARVE2018}, and the minimal requirements of Google Stadia. 
Over this testbed, we generate, encode, transmit, decode, and display a 1920$\times$1080 video, encoded at a bitrate of 6.5\,Mb/s and a frame rate of 30\,FPS, with GoPs of size 10.

\begin{figure*}[!t]
    \centering
\begin{subfigure}[b]{0.49\textwidth}    
\centering
    \includegraphics[width=\textwidth]{ 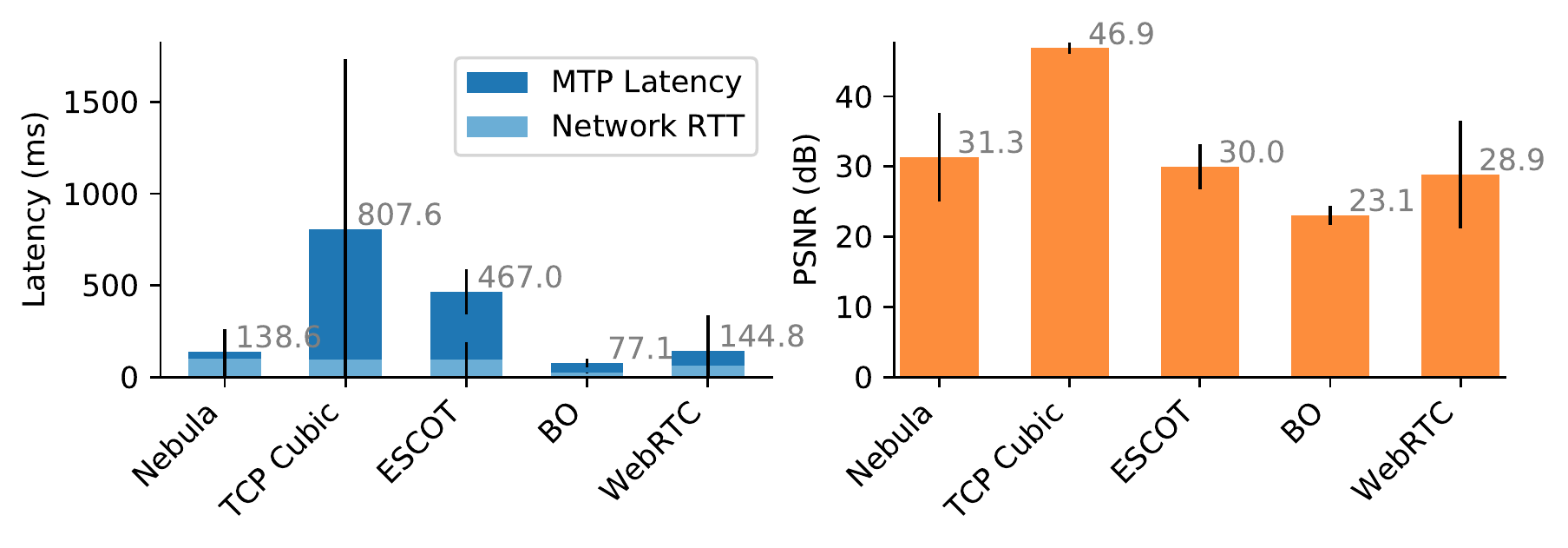}
    \caption{Emulated Network}
    \label{fig:mtp}
    \end{subfigure}\hfill\begin{subfigure}[b]{0.49\textwidth}
    \centering
    \includegraphics[width=\textwidth]{ 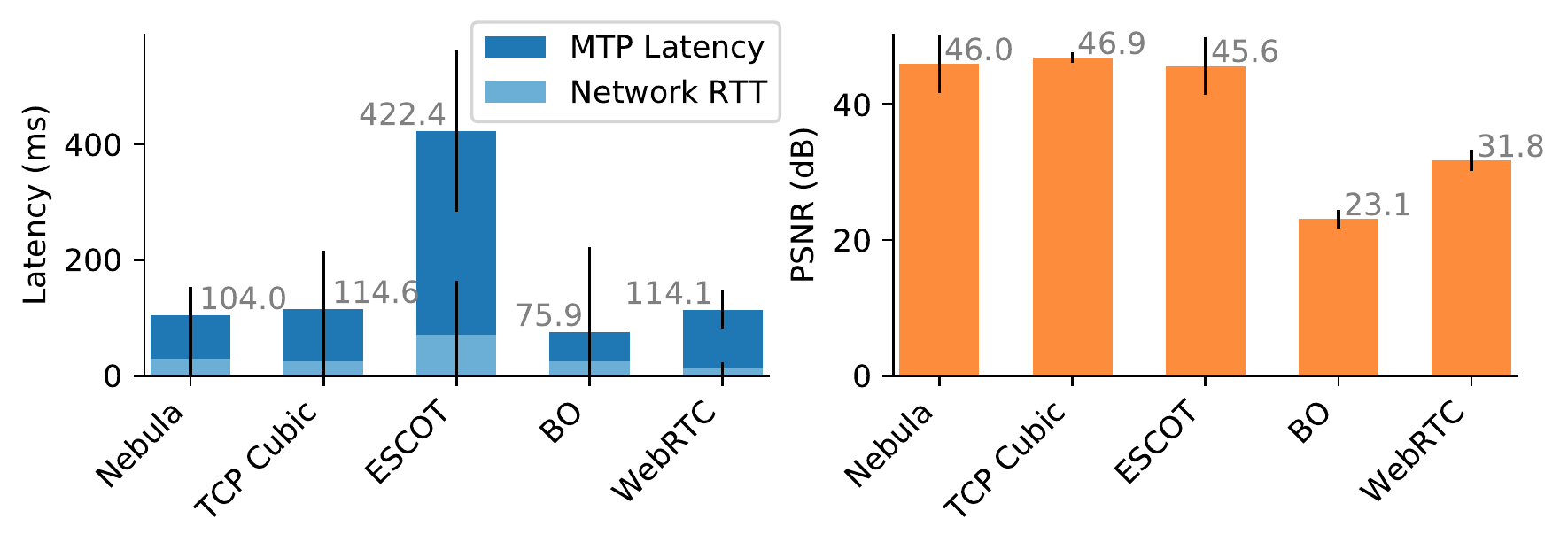}
    \caption{Eduroam WiFi Network}
    \label{fig:mtp_wifi}
    \end{subfigure}
    \vspace{-0.2cm}
    \caption{MTP latency, network RTT, and PSNR on emulated network (a) and  eduroam WiFi network (b). On the emulated network (a), WebRTC and \sysname balance low MTP latency and high visual quality. BO achieves lower MTP latency with considerable visual quality degradation, while TCP Cubic and ESCOT suffer from massive MTP latency.  On the eduroam WiFi (b), all solutions achieve a MTP latency lower than 130\,ms except ESCOT. WebRTC's visual quality collapses due to the high jitter, while TCP Cubic remains stable thanks to the higher bandwidth. In both scenarios, \sysname presents the second best visual quality and the second lowest MTP latency.   } 
    \label{fig:mtp_overall}
\end{figure*}

Figure~\ref{fig:latency} presents the latency decomposition of the pipeline without the network latency. Introducing RLNC adds a marginal delay before (0.8\,ms), and after (1.1\,ms) the transmission. Video encoding and decoding take the most time, 20.8 and 9.1\,ms respectively as the libvpx operates solely on CPU. By leveraging hardware-level video encoding and decoding as it would be the case on mobile devices, we expect to reduce these values to 5-10\,ms\footnote{\url{https://developer.nvidia.com/nvidia-video-codec-sdk}}, and achieve a pipeline latency below 30\,ms. 
\sysname thus does not introduce significant latency compared to typical cloud gaming systems, while bringing loss recovery capabilities in a lossy transmission environment.

\subsection{Emulated Network}
\label{sec:emulatednet}

\begin{figure}[t]
    \centering
    \includegraphics[width=.9\columnwidth]{ 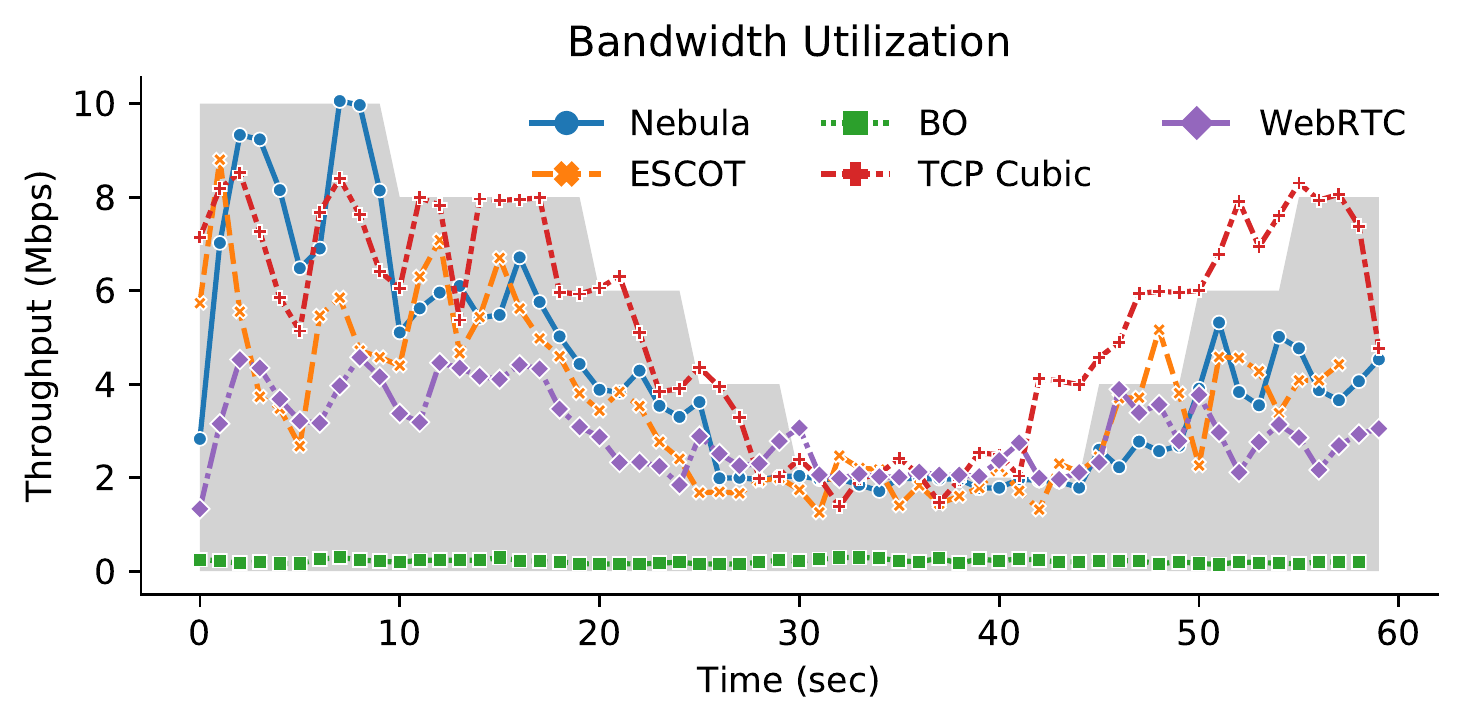}
    \vspace{-0.2cm}
    \caption{Throughput of each connection with variation of the link bandwidth (grey). \sysname uses the bandwidth the most efficiently. ESCOT and TCP Cubic tend to overshoot while WebRTC significantly undershoots. BO barely utilizes the available bandwidth. }
    \label{fig:latencyreliabilityvariable}
\end{figure}

To showcase the loss recovery and rate adaptation properties of the considered solutions, we emulate a wireless network with varying bandwidth over our physical testbed. On this link, we set up a bandwidth ranging between 2\,Mb/s and 10\,Mb/s, changing every 5\,s. Such values allow us to stress the transmission solutions 
 over variable network conditions.  To ensure the reproducibility of the experiments, we generate a  sequence of bandwidth allocation over time represented by the gray filled shape in Figure\ref{fig:latencyreliabilityvariable} and average the results over five runs for each solution. The link round-trip delay and loss probability remain constant over time, standing at 20\,ms and 1\% with probability of successive losses of 25\%, respectively. Over this link, we transmit 60 seconds of a gameplay video, recorded at a resolution of 1080p, and encoded in VP8 in various resolutions by the transmission scheme.

Figure~\ref{fig:mtp_overall}.\subref{fig:mtp} represents the MTP latency, network RTT, and average PSNR for all solutions. Only BO satisfies the requirement of MTP latency below 130\,ms in such constrained environment, while \sysname and WebRTC remain close (138.6\,ms and 144.8\,ms respectively). TCP Cubic collapses under the abrupt bandwidth changes and presents the highest motion-to-photon latency (807.6\,ms), with a high variance caused by the loss retransmission. Many frames do not reach on-time (330\,ms with our pipeline) and are thus dropped by the pipeline. ESCOT also shows high MTP latency (489.8\,ms) due to its GoP-level FEC encoding. As a result, ESCOT requires to receive the entire GoP before decoding, resulting in 330\,ms added latency at 30\,FPS with a GoP of 10. In terms of PSNR, BO can maintain such a low latency by significantly decreasing the video quality. TCP Cubic presents the highest PSNR as it does not integrate a mechanism to vary the video encoding quality. Only \sysname and WebRTC balance latency and video quality, with \sysname presenting slightly higher PSNR and lower MTP latency.


To better understand this phenomenon, we record the received rate at the client's physical interface and present the throughput of each solution in Figure~\ref{fig:latencyreliabilityvariable}. BO strives to eliminate video playback stalls by keeping a reservoir of frames with minimum rate. It thus dramatically underestimates the available bandwidth, between 191 and 542\,Kb/s. WebRTC consistently sends data at a rate much lower than the available bandwidth, while TCP Cubic tends to overshoot due to aggressive congestion window increase. Although \sysname and ESCOT rely on the same bandwidth estimation mechanism, ESCOT unsurprisingly tends to slightly overshoot by transmitting the entire GoP at once. By optimizing bandwidth usage, \sysname maximizes the video quality while minimising latency and losses. 

\subsection{Wireless Network}
\label{sec:wireless}

Following the evaluation on emulated links, we proceed to an experiment on a real-life uncontrolled WiFi network. We perform the experiment on \texttt{eduroam}, a network commonly used by all students of the university. To ensure the statistical significance of the results, we transmit the video from the previous experiment five times per solution and average the results. We first characterize the network by first sending a continuous flow using iPerf over 10 minutes. The network features an average bandwidth of 8.1\,Mb/s (std=5.1\,Mb/s) and latency of 22.7\,ms (std=38.8\,ms). These average values are fairly similar to our emulated network. However, the variability of the available bandwidth is lower while the latency jitter skyrockets, with up to 70\,ms difference between measurement points.


Figure~\ref{fig:mtp_overall}.\ref{fig:mtp_wifi} represents the MTP latency, network RTT, and average PSNR for all solutions. With the more stable bandwidth, all solutions except ESCOT achieve a MTP latency below 130\,ms. Similar to the previous experiment, BO achieves minimal MTP latency, at the cost of a PSNR half of the other solutions. WebRTC collapses under the high variability of the latency, resulting in a low PSNR. ESCOT still presents high MTP latency by transmitting frames by GoP-sized batches. Finally, \sysname and TCP Cubic perform similarly, with low MTP latency and high PSNR.

From the emulated network and uncontrolled WiFi experiments, only \sysname performs on average the best, balancing latency and visual quality consistently. Although TCP's behavior with varying bandwidth and the moderate loss rate is known and expected, we were surprised to notice the collapse of the visual quality of WebRTC in environments with high latency jitter, as is the case in most public wireless and mobile networks. Overall, ESCOT performs poorly due to the GoP size used in the experiments. However, a lower GoP would lead to larger videos for similar quality, and we expect the PSNR to drop with the MTP latency. Finally, BO often underestimates the bandwidth and consistently chooses the lowest quality video flow, leading to a dramatically low PSNR.


\subsection{User Study}

\begin{figure}[t]
    \centering
    \includegraphics[width=.9\linewidth]{ 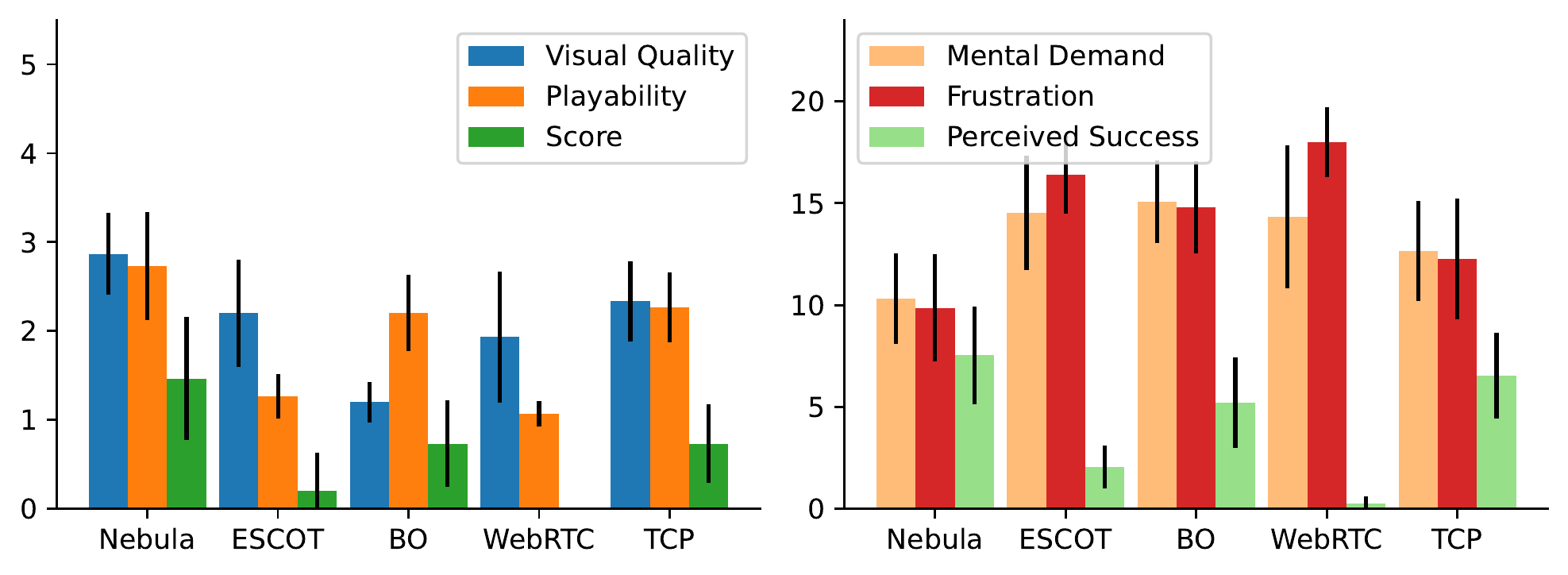}
    \caption{Users' perception of the gaming experience under traditional cloud gaming measures (left) and task load (right). \sysname presents the highest visual quality and playability and the lowest mental demand and frustration of all solutions. Score and perceived success are also the highest with \sysname. WebRTC collapses under the high jitter of the WiFi network, resulting in low playability. }
    \label{fig:userQoE}
\end{figure}

\textbf{Participants and Apparatus:} We perform a user study with 15 participants, aged 20 to 40. The participants are recruited on campus and are primarily students and staff. Most participants play video games at least a few times a month and have some familiarity with First Person Shooter games. The participants play the game openarena in a cloud gaming setting using the prototype system described in Section~\ref{sec:implem} and the transmission techniques defined in Section~\ref{sec:testbed} over the eduroam WiFi network. 

\noindent\textbf{Protocol:} Each participant starts with 5 minutes of free play of the game openarena executed on-device. This phase allows the participants to get familiar with the game and establish a reference of playability and visual quality. During this phase, an operator guides the participants and answers their questions. The participants then play the game for two minutes for each streaming method. The order in which the participants experience each streaming solution follows the balanced latin square design~\cite{williams1949experimental} to avoid learning and order effect.  After each run, the participants fill a short questionnaire on the perceived visual quality and playability on a 5-point Likert scale (1 - bad, 2 - poor, 3 - fair, 4 - good, 5 - excellent), and a simplified version of the NASA TLX~\cite{hart2006nasa} survey considering the perceived mental demand, frustration, and success on a [0-20] scale. We do not disclose the streaming methods used in order not to affect the participants' ratings.

\noindent\textbf{Results:} Figure~\ref{fig:userQoE} presents the results of the user study, with the 95\% confidence intervals as error bars. \sysname results in the highest visual quality and playability, as well as the lowest mental demand and frustration. \sysname also leads to the highest objective (score) and subjective (success) performance indicators. TCP also performs well. However, the inability to reduce the video quality in case of congestion causes transient episodes of high latency that leads the server to drop the frames, resulting in lower visual quality and playability. Similar to Section~\ref{sec:wireless}, BO leads to low latency, explaining the high playability score. However, the video quality drops so low that players have trouble distinguishing what happens on screen, resulting in high mental demand and frustration. WebRTC collapses due to the high jitter, leading to an almost-unplayable experience. Finally, the high latency of GoP-level encoding proves to be highly incompatible with a fast-paced game such as openarena. Overall, \sysname maintains latency consistently low while adapting the visual quality to the best achievable quality given the network conditions, leading to the highest user satisfaction.

\section{Conclusion}
\label{sec:conclusion}

This paper introduced \sysname, an end-to-end framework combining per-frame adaptive FEC with source rate control to provide MCG with low-latency yet error-resilient video streaming capabilities.  
By combining adaptive FEC and source rate adaptation, \sysname minimizes the distortion propagation while maximizing the bandwidth usage, leading to a high end-to-end PSNR without sacrificing latency. 
Our system evaluation shows that \sysname balances low latency and high PSNR under all scenarios, even in highly variable conditions where other solutions (TCP Cubic and WebRTC) collapse.
Under high latency jitter, it significantly outperforms WebRTC, the current industry standard for video transmission in cloud gaming. 
Our user study over a real-life public WiFi network confirms these findings, with \sysname consistently presenting higher visual quality, playability, and user performance while requiring a lower workload than the other solutions.

We aim to focus on multiplayer streaming in our future work, where several devices share a given scene. By generating the video in multiple resolutions, our system allows distributing content to clients connected through variable network conditions. We will also continue the study of \sysname over mobile networks to evaluate the effect of mobility on the frame transmission of video games and the resulting user experience.


\bibliographystyle{ACM-Reference-Format}
\balance
\bibliography{main.bib}

\newpage
\nobalance
\section{Supplementary Materials}
\label{sec:appendix}
\subsection{System Protocol Packets}
\label{sec:nebulapackets}
\begin{minipage}{0.45\textwidth}
    \centering
    \includegraphics[width=0.9\linewidth]{ 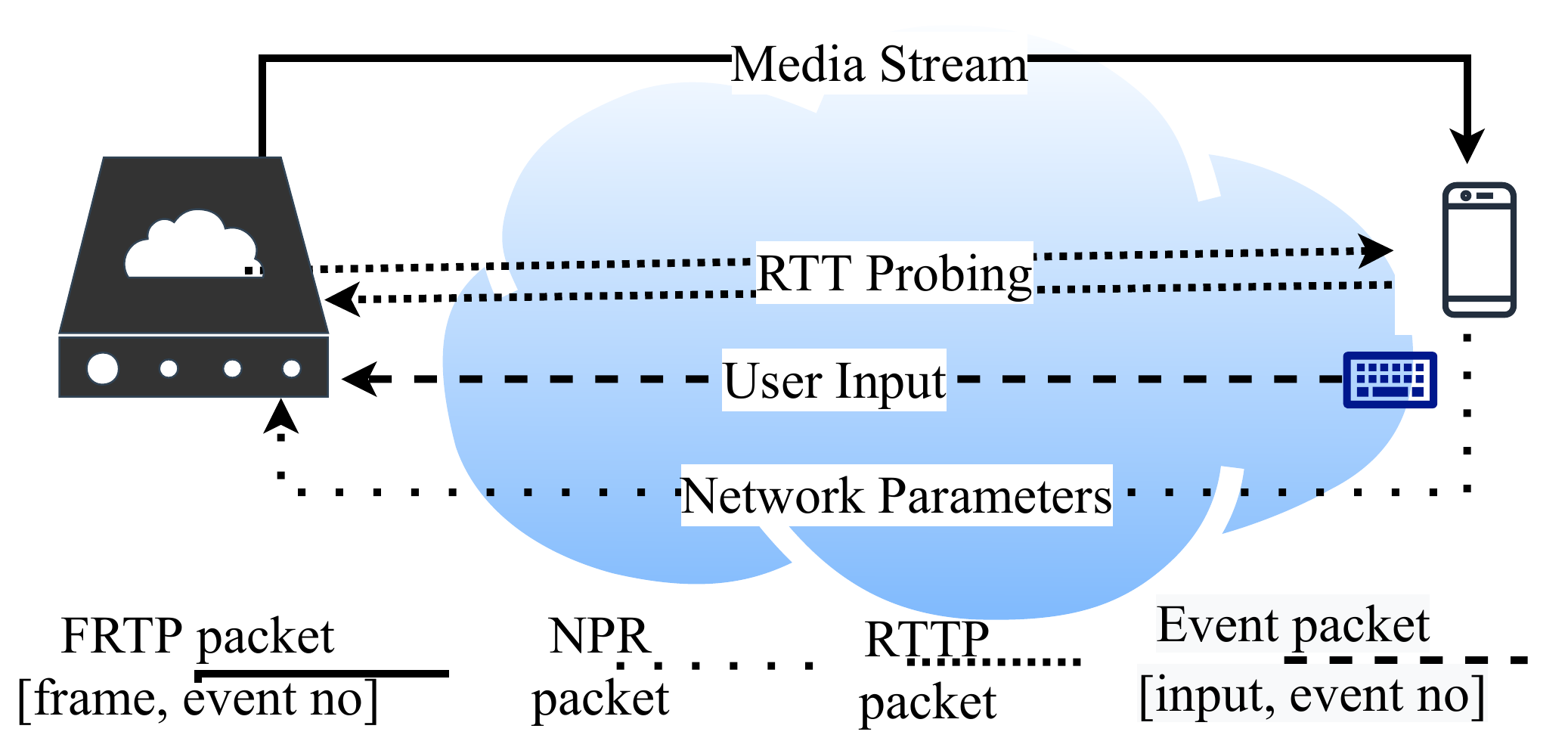}
    \captionof{figure}{Illustration of media delivery and RTT probing on the forward channel and feedback reporting and user input on the reverse channel.}
    \label{fig:nebulapackets}

\vspace{1em}
\raggedright
In this section, we illustrate \sysname's transmission protocol packets as an integral component in the \textbf{prototype system implementation} (see Section~\ref{sec:pipelineimplem}). Figure~\ref{fig:nebulapackets} illustrates the packet type used to deliver the game frames (i.e., FRTP packet), the one used to collect reports about the network conditions (i.e., NPR), the packet type used to probe the presence of congestion (i.e., RTTP  packet), and the one used to send the user input to the server (i.e., Event packet). The server sends packets carrying the encoded game frame's data to the client using FEC Forward Recovery Realtime Transport Protocol (FRTP) packets. FRTP packets contains the event's sequence number if the frame is rendered in response to a user's event. The client reports the experienced network conditions and MTP periodically using the Network Performance/Parameters Report (NPR) packets. It also sends user input as event identifier and number using Event packets. Moreover, the server probes the round trip time periodically using RTT Probing (RTTP) packets. Eventually, the client can compute MTP latency by subtracting the timestamp of sent event from the timestamp of a received frame having a corresponding event sequence number. 
\end{minipage}

\subsection{Heuristic Algorithm Detail}
\begin{figure}[t]
    \centering
    \subfloat[FEC adaptability]
    {\includegraphics[height=2.2cm]{ 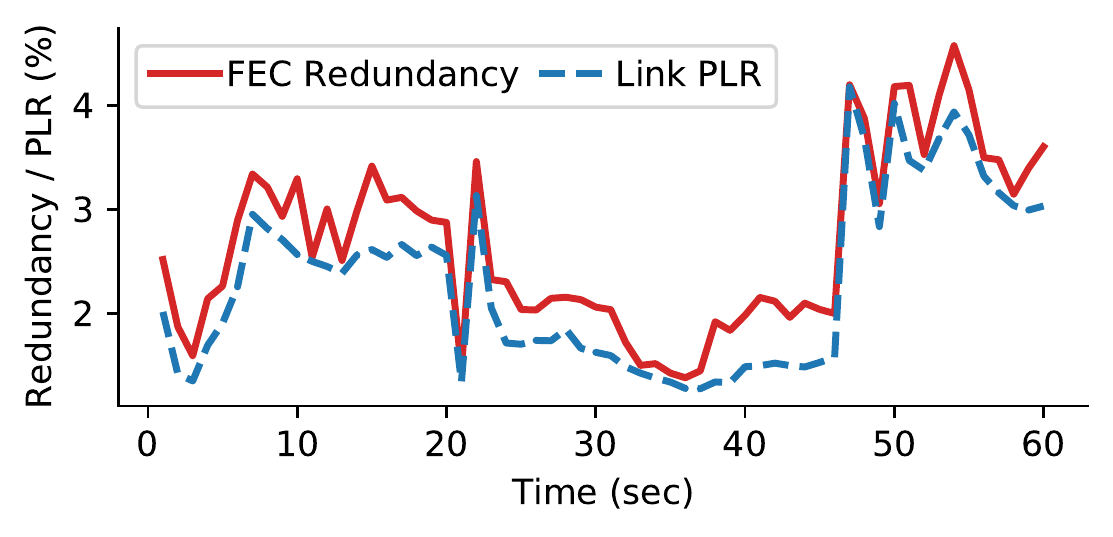} \label{fig:adapt_fec}}
    \subfloat[Rate adaptability]
    {\includegraphics[height=2.2cm]{  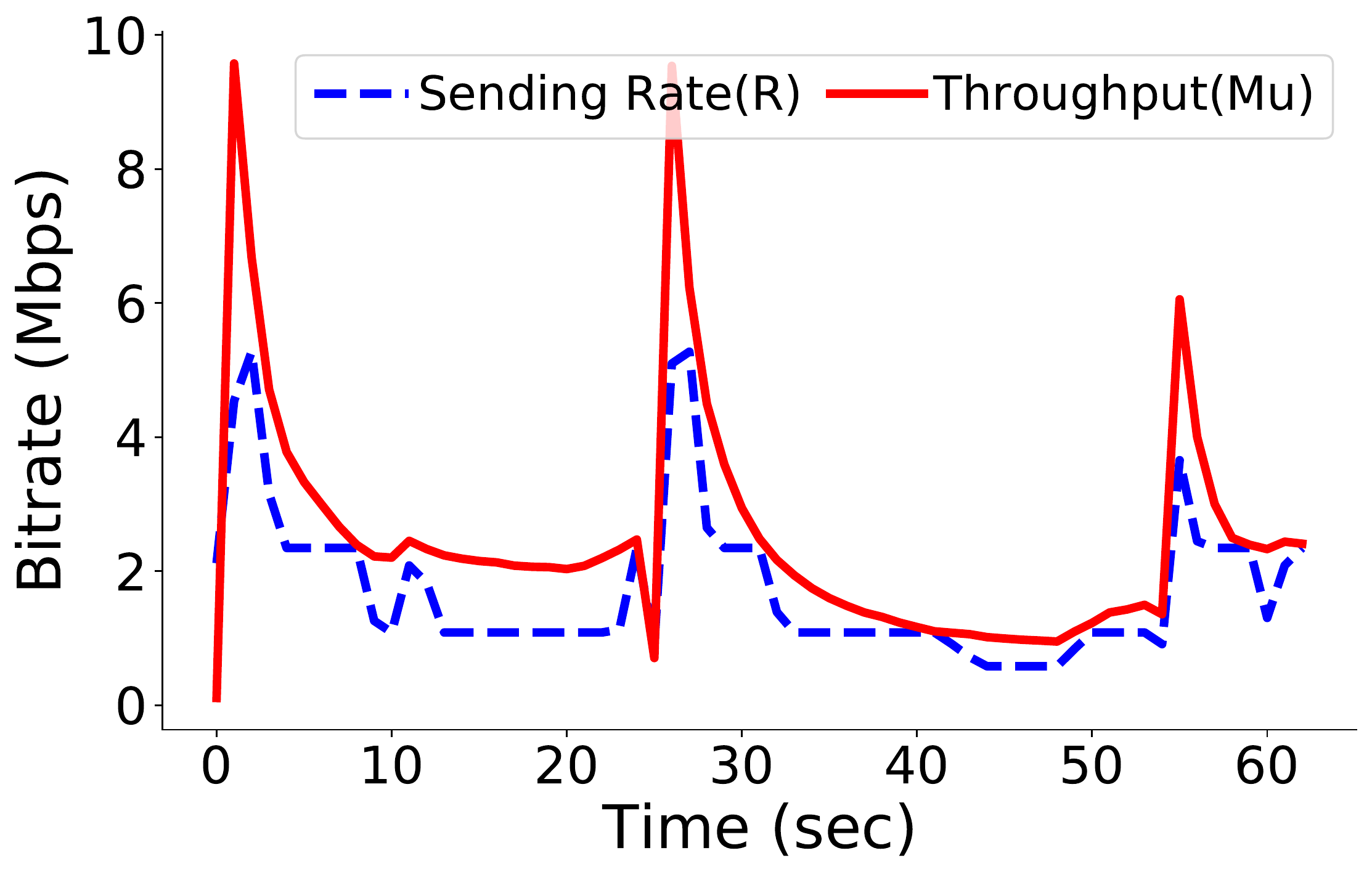}\label{fig:adapt_rate}}
    \caption{Illustration of redundancy and sending rate given a streaming of 1-minute gameplay video over emulated network (see Section~\ref{sec:emulatednet}). (a) FEC redundant information percentage (solid line) vs packet loss rate $PLR$ ($\Pi$) (dashed line) for frames over 40 packets, and (b) sending rate vs network throughput.}
    \label{fig:adapt_sendrate}
\end{figure}

\sysname’s rate and FEC adaptation is based on discrete optimization which follows the heuristic model (see section~\ref{sec:heuristic}). Algorithm~\ref{pseudo_heuristic} presents the pseudo-code of the heuristic approach. It takes as input: the ascending ordered list $list_{R_e}$, the previous encoding rate $last_{R_e}$, the size of GoP frame ${S}_{f_{GoP}}$, and the recent $MTP$ as well as constants, namely GoP length $F$ and end-to-end delay upper-bound $T_d$. It returns the redundancy rate $R_r$ and the video encoding rate $R_e$. After determining the number of packets $k$ and computing the number redundant packets $r$, it computed the redundancy rate $R_r$. Afterwards, it determines the video encoding rate $R_e$ heuristically (lines 8-10) based on network throughput $\mu$ and queuing delay $Q_d$ and according to equation~\ref{eq:sendrate}. Finally, it refines the video encoding rate based on the experienced motion-to-photon latency $MTP$ which ensures not to exceed the latency constraint (i.e., $T_d=130\,ms$) . $level$ in lines 16 or 19 determines the resultant quality level (resolution) and thus the suitable video encoding rate from the list $list_{R_e}$. Given the heuristic algorithm, Figure~\ref{fig:adapt_sendrate} illustrates how adaptive are the redundancy and the total sending rate $R$ as response to packet loss rate and throughput changes. By considering all the frames, however, the redundancy rate show in Figure~\ref{fig:adapt_fec} increases reaching 9\% on average, as described in Section~\ref{sec:overhead}.

\begin{algorithm}[t]
\captionof{algorithm}{Pseudo-code of heuristic algorithm}
\label{pseudo_heuristic}
\begin{algorithmic}[1]
    \State  \textbf{Input:} {$\mu, \beta, Q_d$, $\Pi$, S, $last_{R_e}$, $list_{R_e}$, ${S}_{f_{GoP}}$, MTP, F, $T_d$} 
    \State \textbf{Output:} $R_r, R_e$ 

     \State  $k \gets \frac{{S}_{f_{GoP}}}{S}$ 
    \State Compute $n$ using equation~\ref{eq:fecrate}
    \State  $r \gets n - k$ 
    \State $R_r \gets r. S. F. \beta \times  \frac{8}{ 1024 \times 1024}$
    
    \If{ 1 - $Q_d$ > 0 }
        \For{Each $level_{R_e}$}
            \If{$list_{R_e}$[$level_{R_e}$] + $R_r \geq \mu (1 - Q_d)$}
               \State $level \gets level_{R_e}$ - 1
               \State break
            \EndIf
        \EndFor
    \EndIf
    \If{$list_{R_e}$[$level$] $\geq last_{R_e}$ \&  MTP > $T_d$}
               \State $level \gets level$ - 1
        \EndIf
    \If { 1 - $Q_d\leq$ 0 | $level<0$ }
        \State $level \gets 0$
    \EndIf
   
   \State $R_e$ = $list_{R_e}[level]$
\end{algorithmic}
\end{algorithm}


\begin{table}[t!]
    \centering
        \caption{Baselines considered in the system evaluation}
    \label{tab:baselines}
    \small
\begin{tabular}{p{1.3cm}|p{2.2cm}|p{3cm}}
\hline
\textbf{Protocol} & \textbf{Loss Recovery} & \textbf{Rate Control}  \\
\midrule
TCP Cubic & NACK & Congestion Window \\
\midrule
WebRTC & Hybrid NACK/FEC & GCC\\
\midrule
BO~\cite{10.1145/2619239.2626296,huang2013bo} & N/A & Buffer Occupancy \\
\midrule
ESCOT~\cite{wu2017streaming} & GoP-level FEC & \multirow{2}{*}{Throughput-latency based} \\
\cline{1-2}
\sysname & Frame-level FEC &    \\
\bottomrule
\end{tabular}
\end{table}

\subsection{Baselines}
\label{sec:baselines}
\begin{figure}[!b]
    \centering
    \includegraphics[width=.8\linewidth]{ 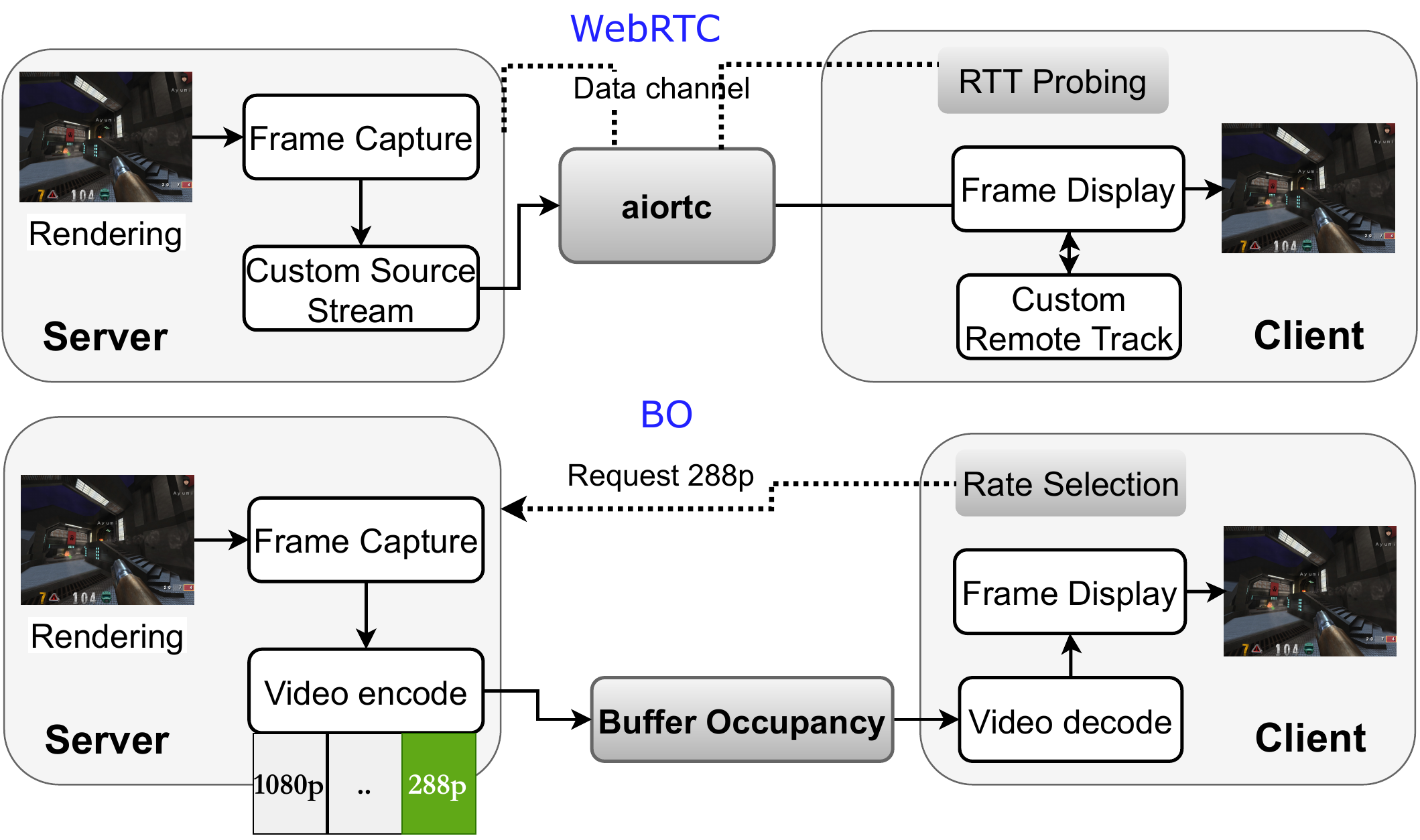}
    \caption{WebRTC's and BO's experimental frameworks.}
    \label{fig:baselinesfw}
\end{figure}

Table~\ref{tab:baselines} illustrated the baselines which are: the network transmission standard (TCP Cubic), the video streaming standard (WebRTC), Buffer Occupancy (BO)-based video streaming~\cite{10.1145/2619239.2626296,huang2013bo}, and GoP-based MCG (ESCOT)~\cite{wu2017streaming}. We integrate both BO and WebRTC to operate as streaming platform for mobile cloud gaming (see Figure~\ref{fig:baselinesfw}). In both, we integrate the captured frames as source stream at the server, and a remote stream track at the client. We enable RTT probing in WebRTC using its data channel and we implement a queue-based playback buffer in BO based on which the Rate Selection component in the client requests the next rate from the server. Such integration of the baselines allows us not only to measure both visual quality and MTP on frame basis but also create reproducible experiments and findings.

\subsection{Overhead}
\label{sec:overhead}
\begin{figure}[!b]
    \centering
    \includegraphics[width=0.9\linewidth]{ 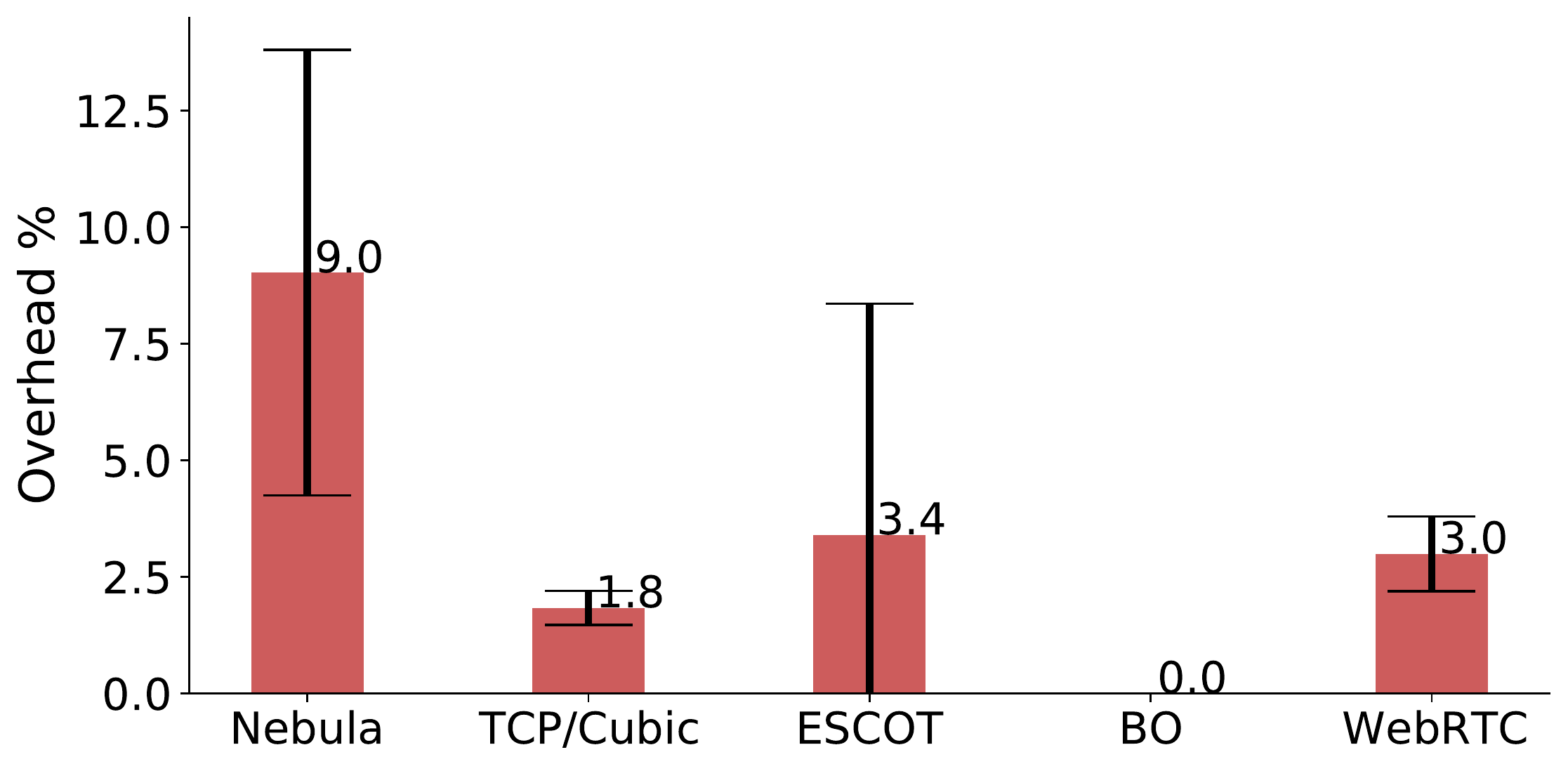}
    \caption{Redundant packet as reliability overhead for each solution. \sysname has the highest, followed by GoP-level FEC (ESCOT), then hybrid FEC/NACK (WebRTC), and finally NACK only (TCP/CUBIC), while BO has none.  }
    \label{fig:overhead}
\end{figure}

Overall, FEC-based approach tends to have a higher reliability overhead compared to ACK-based techniques that requires overhead equals to the packet loss rate. Figure~\ref{fig:overhead} illustrates the reliability overhead of \sysname and the baselines's streaming over the emulated network described in Section~\ref{sec:emulatednet}.

\sysname's satisfactory visual quality is obtained at the expense of redundancy overhead. \sysname presents a redundancy rate of 9\% on average, due to the per-frame encoding and the minimum redundancy. Each frame require a minimum of 1 redundant packet, and P-frames tend to be small (a couple of packets). Besides, unequally protecting the GoP frames leads to extra redundancy.  As such, the overall overhead becomes relatively high, compared to GoP-level FEC coding. However, such overhead is compensated by the better throughput utilisation, minimizing its impact on the transmission.

\end{document}